\documentclass[aps,prl,reprint,groupedaddress,amsmath,amssymb,floatfix]{revtex4-2}
\usepackage[T1]{fontenc}
\usepackage[utf8]{inputenc}
\usepackage{graphicx}
\usepackage{bm}
\usepackage[mathlines]{lineno}
\renewcommand{\linenumberfont}{\normalfont\tiny\sffamily}
\usepackage{hyperref}
\hypersetup{hidelinks,breaklinks=true}
\begin{document}

\title{Field-free Magnetoacoustic Nonreciprocity via Exchange-bias Engineering}

\author{Weiqi Du}

\author{Mingyuan Ma}

\author{Zheng Sun}

\author{Chong Chen}

\author{Shixuan Liang}

\author{Deqi Tao}

\author{Shihai An}

\author{Yifan Fu}

\author{Sulei Fu}

\author{Feng Pan}

\author{Cheng Song}

\email{songcheng@mail.tsinghua.edu.cn}

\affiliation{Key Laboratory of Advanced Materials (MOE), School of Materials Science and Engineering, Tsinghua University, Beijing 100084, China}

\begin{abstract}

Nonreciprocal wave propagation is essential for directional signal processing and isolation. Magnetoacoustic systems are attractive platforms for nonreciprocal devices because of their small footprint and passive operation. However, pronounced nonreciprocity typically requires an external bias field, limiting device miniaturization and integration. Here, we demonstrate field-free magnetoacoustic nonreciprocity in an IrMn/\allowbreak{}FeGaB/\allowbreak{}Cu/\allowbreak{}FeGaB multilayer integrated with a shear-horizontal-wave delay line. Exchange-bias engineering stabilizes a remanent magnetic state with broken time-reversal and spatial-inversion symmetries while tailoring the spin-wave dispersion to achieve resonant magnetoacoustic coupling at zero field. The normalized isolation reaches 21 dB/\allowbreak{}mm, with its sign programmable by the field-sweep history. Further numerical optimization predicts a zero-field normalized isolation of 91 dB/\allowbreak{}mm by increasing the dynamical magnetic asymmetry. Our work establishes exchange-bias engineering as a promising route toward compact RF isolators and circulators and as a general strategy for realizing magnetoacoustic devices without external magnetic fields.

\end{abstract}


\maketitle

% arXiv copy: review line numbering disabled.

\nolinenumbers
% Avoid artificial vertical stretching between paragraphs and around equations.
\raggedbottom
\setlength{\parskip}{0pt}
\setlength{\abovedisplayskip}{6pt}
\setlength{\belowdisplayskip}{6pt}
\setlength{\abovedisplayshortskip}{3pt}
\setlength{\belowdisplayshortskip}{3pt}

% Source Word paragraph 13

Nonreciprocal wave propagation is essential for controlling the directional flow of energy and information, underpinning broad applications in signal processing and communication systems~\cite{ref26}. A central route toward nonreciprocity (NR) is symmetry engineering, namely, breaking the symmetries that relate counterpropagating waves. Nonreciprocity has been widely explored in electromagnetic, mechanical, and spin-wave systems through various mechanisms including magnetic biasing~\cite{ref25,ref27,ref32}, spatiotemporal modulation~\cite{ref24}, nonlinear non-Hermitian effects~\cite{ref22}, among others~\cite{ref23}.

% Early float anchor for Fig. 1.

\begin{figure*}[!t]
\centering
\includegraphics[width=0.76\textwidth]{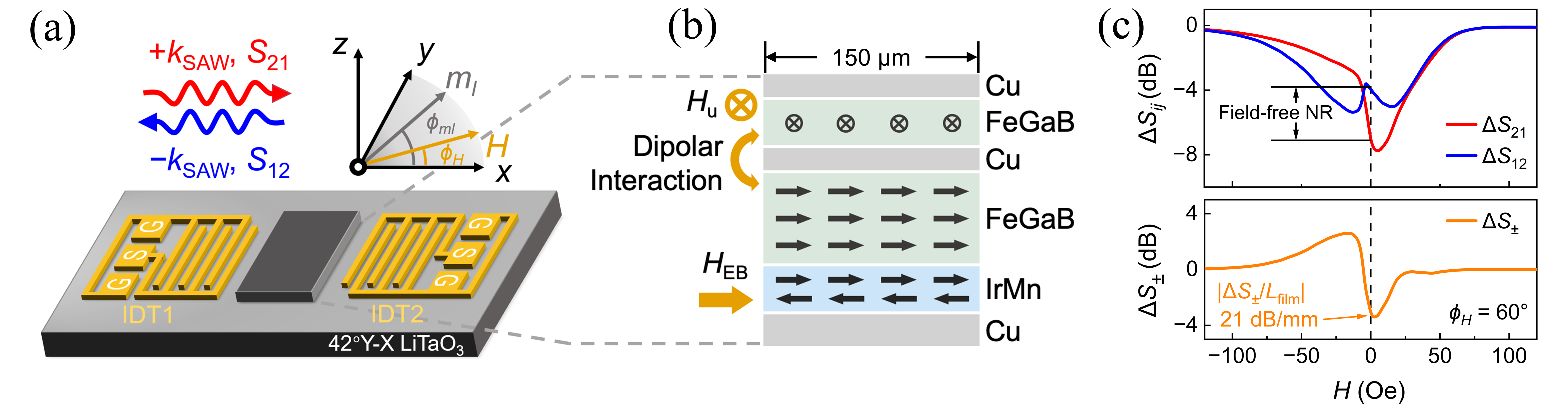}
\caption{Device concept and representative results for field-free nonreciprocal magnetoacoustic waves. (a) Schematic of device structure and the coordinate system. A ferromagnetic multilayer is embedded in a \(42^{\circ}\) Y-X SH-SAW delay line. An in-plane magnetic field \(H\) is applied at an angle \(\phi _{\mathrm{H}}\) with respect to the SAW propagation direction \(+x\). (b) Cross-sectional schematic of the exchange-biased ferromagnetic bilayer. The IrMn layer introduces an exchange-bias field \(H_{\mathrm{EB}}\) in the lower FeGaB layer, while the two FeGaB layers are coupled through the interlayer dipolar interaction. The in-plane uniaxial anisotropy field \(H_{u}\) of the upper FeGaB layer is nearly orthogonal to \(H_{\mathrm{EB}}\). (c) Representative results of field-dependent SAW transmission attenuation \(\Delta S_{ij} (ij = 21,12)\) and its nonreciprocity \(\Delta S_{\pm }\) for \(\phi _{H} = 60^{\circ}\). Pronounced magnetoacoustic nonreciprocity appears around the zero-field point and the normalized isolation reaches 21 dB/\allowbreak{}mm.}
\label{fig:1}
\end{figure*}

% Source Word paragraph 14

Magnetoacoustic systems are particularly attractive for nonreciprocal devices because of their compact footprint and passive operation~\cite{ref39}. A typical magnetoacoustic device consists of a surface acoustic wave (SAW) delay line coupled to a magnetic layer placed in the acoustic propagation path~\cite{ref27,ref28,ref36,ref37}. In such systems, nonreciprocity arises when magnetic states and wave modes break proper time-reversal symmetry (TRS) and spatial-inversion symmetry (SIS). Various mechanisms have been investigated to realize magnetoacoustic nonreciprocity~\cite{ref09,ref12,ref11,ref10,ref14,ref19,ref20,ref17}, among which magnetoelastic coupling is the most commonly exploited and has enabled large isolation~\cite{ref13,ref15,ref21}, nonreciprocal phase shifts~\cite{ref40}, and broadband directional transmission~\cite{ref18}. However, pronounced nonreciprocity generally requires an external bias field, compromising the compactness and integrability of SAW-based devices. This motivates a built-in approach that breaks both TRS and SIS~\cite{ref09,ref10} while preserving resonant magnetoacoustic coupling at zero field. It is challenging to achieve because unstable multidomain magnetic states and resonance detuning at low fields strongly suppress the nonreciprocal response.

% Source Word paragraph 15

Exchange bias (EB) provides a natural route toward this goal~\cite{ref01,ref02,ref03}. Originating from interfacial exchange coupling between a ferromagnet and an antiferromagnet, EB introduces a unidirectional magnetic anisotropy that acts as a built-in Zeeman-like effective field (\(H_{\mathrm{EB}}\)). This effect has long been used to pin ferromagnetic reference layers and stabilize magnetic configurations in magnetic tunnel junctions~\cite{ref08}, and has also been exploited to provide built-in symmetry breaking for field-free deterministic magnetization switching~\cite{ref05,ref06}. Beyond such qualitative control, EB also offers a quantitative means of engineering the magnetic energy landscape~\cite{ref04}. This tunability allows \(H_{\mathrm{EB}}\) to tailor both the magnetic hysteresis and spin-wave dispersion, enabling specific magnetic responses, such as ferromagnetic resonance, to be precisely positioned at zero applied field. These capabilities of EB closely match the key requirements for realizing field-free magnetoacoustic nonreciprocity. Moreover, the interfacial nature of EB makes it especially well suited to thin-film magnetoacoustic devices, where EB can exert substantial control over the properties of nanometer-thick ferromagnetic films.

% Source Word paragraph 16

In this Letter, we demonstrate field-free magnetoacoustic nonreciprocity in an exchange-biased IrMn/\allowbreak{}FeGaB/\allowbreak{}Cu/\allowbreak{}FeGaB heterostructure integrated with a shear-horizontal-wave (SH-wave) delay line. EB stabilizes a symmetry-broken remanent magnetic state and tunes the spin-wave spectrum toward the SAW frequency, giving rise to pronounced nonreciprocity at zero field. We first present the stepwise design evolution to achieve field-free nonreciprocity. Then, we characterize the angle-dependent behavior, showing a stable zero-field response whose sign can be programmed by the field-sweep history. Finally, numerical calculations reveal how exchange-bias engineering enhances the dynamical asymmetry of the ferromagnetic bilayer, increasing the \(\pm k\) spin-wave splitting and thereby strengthening the zero-field nonreciprocity. Beyond demonstrating field-free magnetoacoustic nonreciprocity itself, our approach provides a general strategy for designing various field-free magnetoacoustic devices~\cite{ref34,ref35}.

% Source Word paragraph 19

Figure~\ref{fig:1}(a) introduces the device geometry designed for realizing field-free magnetoacoustic nonreciprocity. The device consists of a \(42^{\circ}\) Y-cut X-propagating LiTaO\textsubscript{3} SH-wave delay line and a magnetic multilayer placed between two interdigital transducers (IDTs). A vector network analyzer is connected to the two IDTs to measure the SAW transmission parameters, \(S_{21}\) (\(+k_{\mathrm{SAW}}\)) and \(S_{12}\) (\(- k_{\mathrm{SAW}}\)). The IDTs generate and detect SH-waves through the (inverse) piezoelectric effect, while the waves propagate along \(\pm x\) and interact with the magnetic multilayer. In our optimized device, the SAW wavelength \(\lambda _{\mathrm{SAW}}\) is 1.4 \(\mu\mathrm{m}\), with a center frequency \(f_{0}\) of 2.89 GHz. The external magnetic field is applied in-plane at an angle \(\phi _{H}\) with respect to \(+x\) axis. As shown in Fig.~\ref{fig:1}(b), the magnetic structure consists of IrMn(10)/\allowbreak{}FeGaB(20)/\allowbreak{}Cu(4)/\allowbreak{}FeGaB(10) (in nanometers) sandwiched by two Cu layers, and is field-cooled along the \(\phi _{\mathrm{EB}}\) direction (See Supplemental Material Secs. I \& II for methods and delay line properties~\cite{ref41}). Through interfacial exchange coupling, the antiferromagnetic IrMn pins the magnetization of the lower FeGaB layer and introduces an exchange-bias field \(H_{\mathrm{EB}}\) along \(\phi _{\mathrm{EB}}\). The Cu spacer suppresses direct interlayer exchange while retaining long-range dipolar coupling.

% Source Word paragraph 20

The SH-wave supported by \(42^{\circ}\) Y-X LiTaO\textsubscript{3} is dominated by the \(u_{y}\) displacement component and is almost linearly polarized (See Supplemental Material Sec. III for finite element simulations). It drives magnetization precession in the two FeGaB layers through an RF magnetoelastic effective field \(\boldsymbol{h}_{i}\)~\cite{ref16,ref29,ref30,ref31,ref33}
\begin{equation}
h_i\propto\epsilon_{xy,i}\cos(2\phi_{m,i}),
\label{eq:magnetoelastic-field}
\end{equation}
where \(\epsilon_{xy,i}\) and \(\phi_{m,i}\) are the shear strain and magnetization angle of layer \(i\) (\(i = \mathrm{L, U}\)). Acoustically driven ferromagnetic resonance (ADFMR) occurs when the field-tuned spin-wave eigenfrequency matches the SAW frequency, resulting in pronounced SAW attenuation~\cite{ref27,ref28,ref38}. Moreover, the negligible phononic angular momentum of this linearly polarized SH mode largely suppresses helicity-mismatch-induced nonreciprocity. Therefore, our observed nonreciprocity can be primarily attributed to exchange bias engineering of nonreciprocal spin waves.

% Source Word paragraph 23

Figure~\ref{fig:1}(c) presents a representative field-dependent magnetoacoustic response measured at \(\phi _{H} = 60^{\circ}\). The background-corrected SAW transmission attenuations for the opposite propagating directions, \(\Delta S_{ij} (ij = 21,12)\), are plotted as functions of the magnetic field \(H\). Their difference defines the magnetoacoustic nonreciprocity, \(\Delta S_{\pm } = \Delta S_{21}  -  \Delta S_{12}\). The ADFMR response is concentrated in the vicinity of zero field, while \(\Delta S_{21}\) and \(\Delta S_{12}\) exhibit pronounced nonreciprocity. \(\Delta S_{12}\) shows two split resonance dips, whereas \(\Delta S_{21}\) exhibits a single, stronger resonance dip. The directional contrast originates from the nonreciprocal spin-wave spectrum of the dipolar-coupled asymmetric bilayer: the \(+k\) and \(- k\) spin-wave branches are differently detuned from the fixed SAW frequency, producing distinct resonant magnetoacoustic absorption. The nonreciprocal response reaches its maximum almost exactly at zero field, directly demonstrating pronounced field-free magnetoacoustic nonreciprocity. The normalized isolation \(|\Delta S_{\pm }\)/\allowbreak{}\(L_{\mathrm{film}}|\) at zero field reaches 21 dB/\allowbreak{}mm, where \(L_{\mathrm{film}} = 150\,\mathrm{\mu m}\) denotes the length of magnetic film along the SAW propagation direction.

% Early float anchor for Fig. 2.

\begin{figure}[!t]
\centering
\includegraphics[width=0.95\columnwidth]{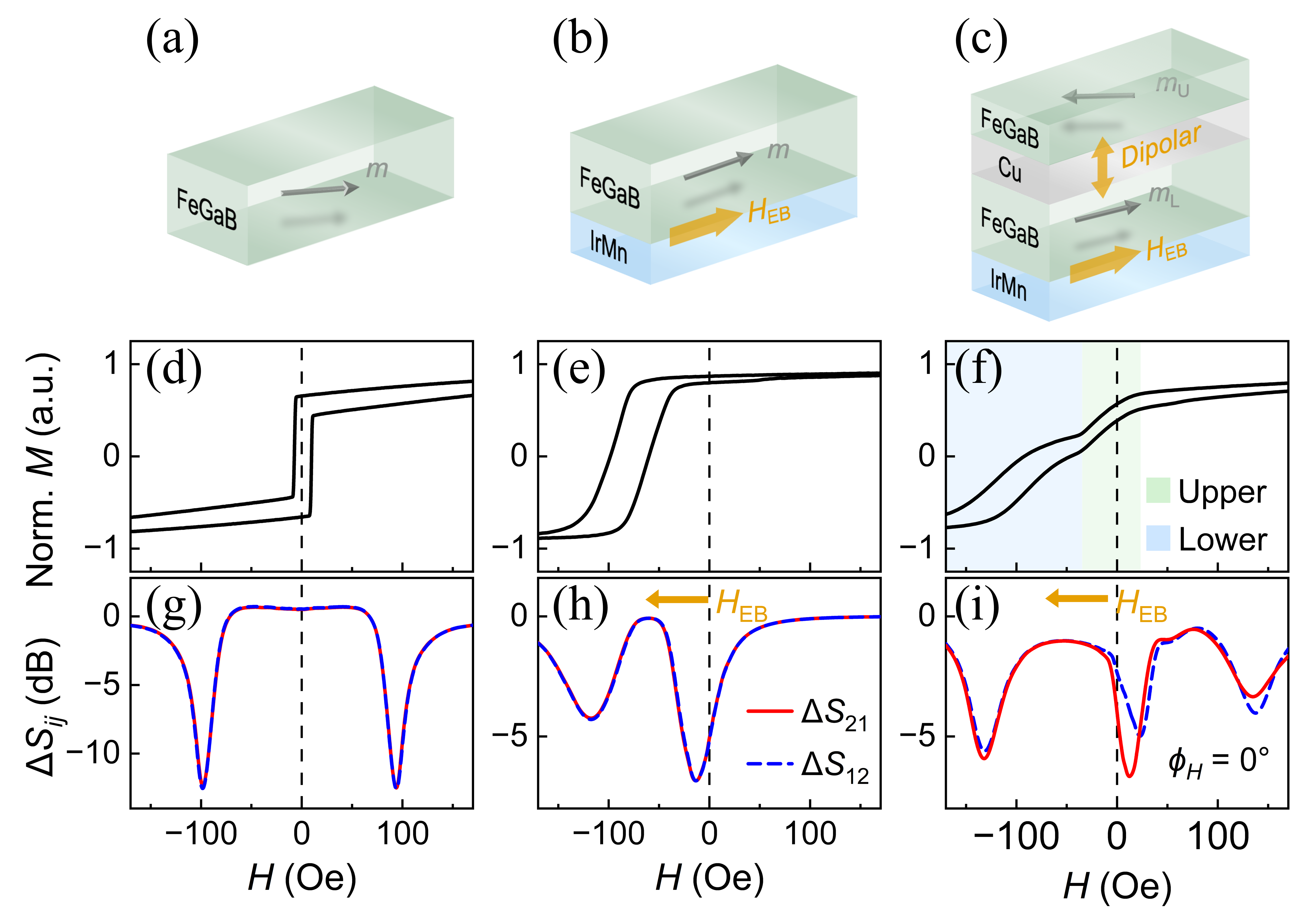}
\caption{Exchange bias engineering to realize field-free magnetoacoustic nonreciprocity. (a--c) Schematics of the magnetic structures embedded in SH-wave delay lines, showing the stepwise design evolution toward field-free ADFMR and field-free nonreciprocity: (a) single FeGaB, (b) exchange-biased IrMn/\allowbreak{}FeGaB, (c) exchange-biased IrMn/\allowbreak{}FeGaB/\allowbreak{}Cu/\allowbreak{}FeGaB with dipolar coupling between the two ferromagnetic layers. (d-f) Corresponding normalized magnetic hysteresis loops measured at \(\phi _{H} = 0^{\circ}\). The shaded regions in (f) indicate the minor magnetic hysteresis loops of lower-FeGaB (blue) and upper-FeGaB (green). (g-i) Field dependence of SAW transmission attenuation \(\Delta S_{ij} (ij = 21,12)\) for the corresponding structures. Pronounced nonreciprocity appears only in the exchange-biased dipolar-coupled bilayer.}
\label{fig:2}
\end{figure}

% Source Word paragraph 28

Next, we will demonstrate the step-by-step process by which field-free nonreciprocity is achieved. We progressively modify the magnetic structure while keeping the same SH-wave delay line geometry, as schematically illustrated in Figs.~\ref{fig:2}(a)-2(c). The normalized magnetic hysteresis loops and the field-dependent SAW transmission attenuations, \(\Delta S_{21}\), \(\Delta S_{12}\), are all measured at \(\phi _{H} = 0^{\circ}\). The SAW frequency is fixed around 2.89 GHz. Starting from a single FeGaB layer, the hysteresis loop shows an easy axis along the \(x\) direction (Fig.~\ref{fig:2}(d)), which is induced by the in-situ magnetic field during film growth. Figure~\ref{fig:2}(g) presents \(\Delta S_{ij} (ij = 21,12)\) as a function of external field. Pronounced SAW attenuation appears when ADFMR occurs, and the resonance fields are symmetric about zero field. The nearly overlapping \(\Delta S_{21}\) and \(\Delta S_{12}\) yield negligible nonreciprocity \(\Delta S_{\pm }\).

% Source Word paragraph 29

To shift the ADFMR condition toward zero field, an antiferromagnetic IrMn layer is introduced underneath FeGaB, and the whole device is field-cooled along \(\phi _{\mathrm{EB}} = 0^{\circ}\). Since the FeGaB layer is sufficiently thin, the EB effect can be treated as a Zeeman-like field \(H_{\mathrm{EB}}\) acting uniformly on the lower-FeGaB layer along \(\phi _{\mathrm{EB}}\). In Fig.~\ref{fig:2}(e), the hysteresis loop of the IrMn/\allowbreak{}FeGaB bilayer is shifted almost entirely into the negative-field region. Consequently, the field-dependent SAW attenuation curves are shifted along \(\phi _{\mathrm{EB}} = \phi _{\mathrm{H}} = 0^{\circ}\), as shown in Fig.~\ref{fig:2}(h), and one of the ADFMR resonances is brought to zero field (See Supplemental Material Sec. IV for the optimization of EB~\cite{ref41}). Still, \(\Delta S_{21}\) and \(\Delta S_{12}\) remain nearly identical, indicating that EB itself does not introduce appreciable nonreciprocity because only TRS is broken at zero field.

% Source Word paragraph 30

Admittedly, integrating such exchange-biased IrMn/\allowbreak{}FeGaB layer with an SIS-broken Rayleigh-wave delay line could also enable zero-field nonreciprocity via helicity mismatch effect, but the resulting normalized isolation is typically small~\cite{ref09}. We therefore introduce a second FeGaB layer separated by a Cu spacer and field-cool the bilayer at \(\phi _{\mathrm{EB}}\) slightly larger than \(0^{\circ}\), further breaking SIS in the magnetic subsystem and forming the final structure shown in Fig.~\ref{fig:2}(c). The hysteresis curve in Fig.~\ref{fig:2}(f) develops two minor loops, corresponding to the exchange-biased lower-FeGaB (blue) and upper-FeGaB (green) layers. The substrate-induced thermal stress rotates the easy axis of the upper FeGaB layer from \(\phi _{u\mathrm{,U}} = 0^{\circ}\) to \(90^{\circ}\) during annealing, whereas that of the lower FeGaB follows \(\phi _{u\mathrm{,L}} = \phi _{\mathrm{EB}}\) due to strong exchange-bias pinning (See Supplemental Material Sec.V for hysteresis loop at \(\phi _{H} = 90^{\circ}\)~\cite{ref41}). The noncollinear easy axes lead to a noncollinear magnetic state in the low-field region, simultaneously breaking SIS and TRS. Consequently, the ADFMR peak near zero field is highly nonreciprocal in Fig.~\ref{fig:2}(i), realizing field-free magnetoacoustic nonreciprocity here. This pronounced response is enabled by the dual role of EB: stabilizing a symmetry-broken magnetic state at zero field and tailoring the spin-wave dispersion to satisfy the resonant magnetoacoustic coupling condition. The other two peaks, appearing at higher fields, mainly originate from the upper FeGaB layer and are nearly symmetric about zero field (see Supplemental Material Sec.~VI for VNA-FMR spectra~\cite{ref41}).

% Source Word paragraph 35

\begin{figure}[!t]
\centering
\includegraphics[width=0.98\columnwidth]{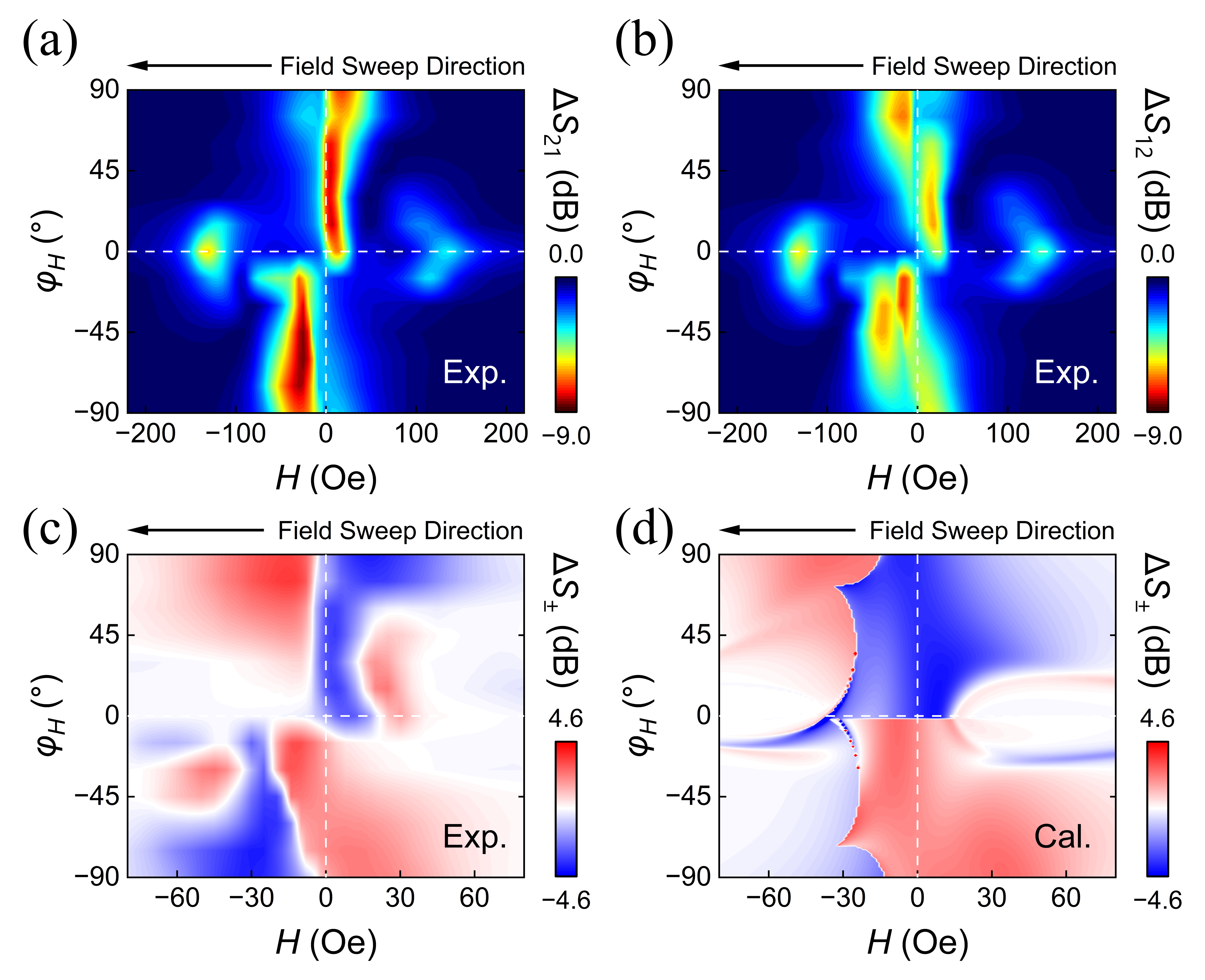}
\caption{Field- and angle-dependence of magnetoacoustic nonreciprocity. (a,b) Experimental results of the SAW transmission attenuation (a) \(\Delta S_{21}\) and (b) \(\Delta S_{12}\) as functions of the field \(H\) and its in-plane angle \(\phi _{H}\). The field is swept from positive to negative. (c,d) Experimental (c) and calculated (d) results of magnetoacoustic nonreciprocity \(\Delta S_{\pm }\) as functions of \(H\) and \(\phi _{H}\).}
\label{fig:3}
\end{figure}

We further studied the angle-dependence of the nonreciprocal magnetoacoustic response in our device. Figures~\ref{fig:3}(a) and 3(b) present the experimentally measured SAW attenuations \(\Delta S_{21}\) and \(\Delta S_{12}\), respectively, as functions of field \(H\) and its in-plane angle \(\phi _{H}\). The field is swept from positive to negative values. In both \(\Delta S_{21}\) and \(\Delta S_{12}\), three prominent resonance regions can be identified. Two nearly reciprocal ADFMR branches appear around the \(\phi _{H} = 0^{\circ}\) axis, consistent with the characteristic angular dependence of SH-wave-driven FMR. In contrast, a much stronger resonance emerges near zero field, and SAW attenuations are highly nonreciprocal here. The resulting magnetoacoustic nonreciprocity \(\Delta S_{\pm }\) is presented in Fig.~\ref{fig:3}(c), with the field range narrowed to \(\pm\)80 Oe. Pronounced nonreciprocity persists at zero field over the entire angular range, with its sign remaining unchanged within approximately 15 Oe around zero, demonstrating a robust field-free response. It is noteworthy that the sign of the zero-field nonreciprocity depends on the field-sweep angle, as presented in Fig.~\ref{fig:3}(c). For \(0^{\circ}  \le  \phi _{H} < 90^{\circ}\), \(\Delta S_{\pm }\) is negative, whereas it becomes positive for \(-  90^{\circ} < \phi _{H} < 0^{\circ}\). This sign reversal mainly originates from the hysteresis of the upper FeGaB layer. Different field-sweep angles \(\phi _{H}\) stabilize different remanent magnetic states in the upper FeGaB layer at zero field, thereby reversing the sign of the nonreciprocal response. In other words, the field-free nonreciprocity in our device is sign-programmable through the field-sweep history. The boundary between the two nonreciprocal polarities is slightly shifted away from \(\phi _{H} = 0^{\circ}\), reflecting the small positive offset of the actual exchange-bias direction \(\phi _{\mathrm{EB}}\) introduced during field-cooling. A control IrMn/\allowbreak{}FeGaB device exhibits negligible nonreciprocity over the same angular range (see Supplemental Material Sec. VII~\cite{ref41}), excluding a potential contribution from exchange-spring effects~\cite{ref07}.

% Fig. 3: delayed until after the experimental discussion.

% Source Word paragraph 36

To further substantiate this interpretation, we developed a theoretical model~\cite{ref28,ref21,ref30} and numerically calculated SAW attenuations in our device (See Supplemental Material Sec. VIII~\cite{ref41}). We calculated \(\Delta S_{\pm}\) using fitted acoustic and magnetic parameters that closely agree with independently measured values, as presented in Fig.~\ref{fig:3}(d). Overall, the calculated \(\Delta S_{\pm }\) map agrees well with the experimental result. Importantly, the calculated zero-field nonreciprocity shows the same sign reversal with \(\phi _{H}\) as observed experimentally, further supporting that the field-free nonreciprocity is governed by the magnetic configuration and established by the field-sweep history. The calculated stability window for field-free nonreciprocity is broader (about \(\pm\)30 Oe) than that observed experimentally. It is mainly attributed to the single-domain approximation used in the calculation, so the upper-FeGaB magnetization remains stable until the field approaches its anisotropy field, \(H_{u\mathrm{,U}}  \approx\) 30 Oe. However, in experiments, domain-wall-mediated magnetization reversal can occur at lower fields, leading to a narrower stability window.

% Fig. 4: right-column top of page 4, alongside Fig. 3.
\begin{figure}[!t]
\centering
\includegraphics[width=0.98\columnwidth]{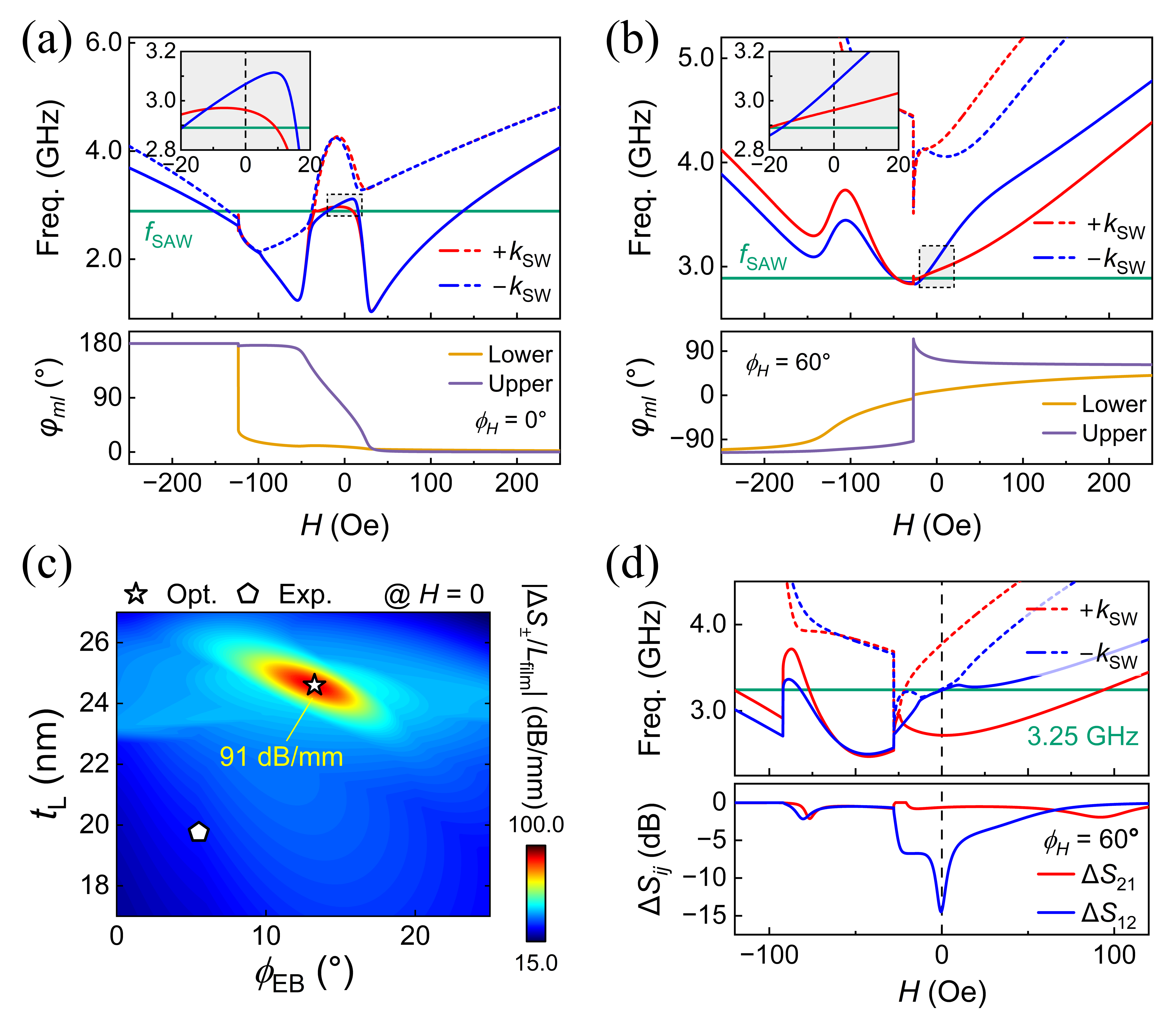}
\caption{Calculated spin-wave dispersions and optimization of field-free magnetoacoustic nonreciprocity. (a,b) Calculated field-dependent spin-wave eigenfrequencies with \(+k_{\mathrm{SW}}\) and \(- k_{\mathrm{SW}}\) at (a) \(\phi _{H} = 0^{\circ}\) and (b) \(\phi _{H} = 60^{\circ}\). The solid and dashed curves denote the two spin-wave modes. Lower panels show the calculated equilibrium magnetization direction \(\phi _{m,i}\) of two FeGaB layers as functions of field.  (c) Calculated zero-field normalized isolation \(\Delta S_{\pm }\)/\allowbreak{}\(L_{\mathrm{film}}\) as functions of \(\phi _{\mathrm{EB}}\) and lower-layer thickness \(t_{\mathrm{L}}\), under fixed \(t_{\mathrm{L}} + t_{\mathrm{U}}\) and \(t_{\mathrm{L}} \cdot H_{\mathrm{EB}}\). The star marks the optimized zero-field \(\Delta S_{\pm }\)/\allowbreak{}\(L_{\mathrm{film}}\) of 91 dB/\allowbreak{}mm. (d) Calculated spin-wave dispersions and corresponding SAW attenuations \(\Delta S_{21}\), \(\Delta S_{12}\) for the optimized structure at \(\phi _{\mathrm{H}} = 60^{\circ}\).}
\label{fig:4}
\end{figure}

% Source Word paragraph 42

Finally, to clarify the role of EB in realizing field-free nonreciprocity, we calculated the spin-wave dispersion and the equilibrium magnetization angles for two representative field-sweep directions, \(\phi _{H} = 0^{\circ}\) (Fig.~\ref{fig:4}(a)) and \(60^{\circ}\) (Fig.~\ref{fig:4}(b)). The SAW frequency, 2.89 GHz, is indicated by the green horizontal line. When \(\phi _{H} = 0^{\circ}\), the spin-wave dispersion of \(+k\) and \(- k\) is nearly degenerate except near the zero-field point. At zero field, the \(+k\) spin-wave branch is much closer to the SAW dispersion, resulting in stronger absorption of the \(+k\)-propagating SAW, consistent with our experimental results. In contrast, at \(\phi _{H} = 60^{\circ}\), the spin-wave dispersions for \(+k\) and \(- k\) become strongly nondegenerate. Combined with the calculated equilibrium magnetization angles, we found that the spin-wave dispersion is nearly degenerate when the two magnetizations are parallel or antiparallel (|\(\phi _{m\mathrm{,L}} - \phi _{m\mathrm{,U}}\)| = \(0^{\circ}\) or \(180^{\circ}\)), whereas pronounced spin-wave nonreciprocity emerges in the noncollinear magnetic state. The noncollinear magnetic state near zero field originates from the different preferred magnetization directions of the two FeGaB layers: exchange bias defines the magnetic anisotropy of the lower-FeGaB layer, whereas substrate-induced thermal stress establishes a different uniaxial anisotropy in the upper-FeGaB, namely \(\phi _{u\mathrm{,L}}  \ne  \phi _{u\mathrm{,U}}\). As discussed above, our SH-wave is almost linearly polarized, so the noncollinear magnetic configuration provides a critical source of dynamical magnetic asymmetry and SIS breaking in our system. It greatly contributes to \(k\)-odd terms in the interlayer dipolar-coupling matrix and hence to nonreciprocal spin-wave dispersion~\cite{ref21}.

% Source Word paragraph 43

Guided by this mechanism, we further explore how to enhance the field-free magnetoacoustic nonreciprocity via exchange-bias engineering. Figure~\ref{fig:4}(c) maps the calculated zero-field normalized isolation \(|\Delta S_{\pm }\)/\allowbreak{}\(L_{\mathrm{film}}|\) as functions of exchange-bias angle \(\phi _{\mathrm{EB}}\) and lower-FeGaB thickness \(t_{\mathrm{L}}\). The total FeGaB thickness is fixed at \(t_{\mathrm{L}} + t_{\mathrm{U}}\) = 30 nm so that the volume of ferromagnets is kept the same as the current experimental configuration, while \(t_{\mathrm{L}} \cdot H_{\mathrm{EB}}\) = 1650 nm\(\cdot\)Oe is kept constant to account for the inverse thickness dependence of exchange bias. All other magnetic parameters are kept the same as those in Fig.~\ref{fig:3}(d). A broad parameter region with strongly enhanced zero-field isolation emerges. By tuning the exchange-bias direction \(\phi _{\mathrm{EB}}\) and the thickness ratio \(t_{\mathrm{L}}\)/\allowbreak{}\(t_{\mathrm{U}}\), the dynamical asymmetry between the exchange-biased and free FeGaB layers can be further enhanced, leading to a larger splitting between the \(+k\) and \(- k\) spin-wave branches and, consequently, stronger magnetoacoustic nonreciprocity. Relaxing the constraints mentioned above allows larger zero-field nonreciprocity to be achieved. 

% Source Word paragraph 44

The maximum calculated zero-field \(|\Delta S_{\pm }\)/\allowbreak{}\(L_{\mathrm{film}}|\) reaches  91 dB/\allowbreak{}mm at \(\phi _{\mathrm{EB}} = 13.25^{\circ}\) and \(t_{\mathrm{L}}\) = 24.6 nm, as marked by the star in Fig.~\ref{fig:4}(c). Figure~\ref{fig:4}(d) shows the corresponding optimized response at \(\phi _{H} = 60^{\circ}\). Around zero field, \(\Delta S_{12}\) exhibits pronounced resonant attenuation, whereas \(\Delta S_{21}\) remains weak, yielding large nonreciprocity with low magnetoacoustic insertion loss of only 0.6 dB. The nonreciprocal response remains pronounced within approximately \(\pm\)15 Oe around zero field. Correspondingly, the \(+k\) and \(- k\) spin-wave branches are strongly nondegenerate near zero field, and the \(- k\) branch is selectively brought into resonance with \(f_{\mathrm{SAW}} = 3.25\,\mathrm{GHz}\). These results establish a general design principle that exchange-bias engineering can simultaneously stabilize the desired zero-field magnetic state and selectively tune one counterpropagating spin-wave branch into acoustic resonance, thereby maximizing field-free nonreciprocity.

% Source Word paragraph 49

In conclusion, we demonstrate field-free magnetoacoustic nonreciprocity in an exchange-biased IrMn/\allowbreak{}FeGaB/\allowbreak{}Cu/\allowbreak{}FeGaB heterostructure integrated with an SH-wave delay line. Exchange bias stabilizes a remanent magnetic state with broken TRS and SIS, and tunes the spin-wave spectrum to enable resonant magnetoacoustic coupling at zero field. The normalized isolation reaches 21 dB/\allowbreak{}mm at zero field and can be sign-programmed by the field-sweep history. Our numerical calculations reproduce the main experimental features and reveal a noncollinear remanent magnetic configuration at zero field, arising from the noncollinear easy axes of the two FeGaB layers. Further enhancing the dynamical asymmetry between the two FeGaB layers can increase the normalized zero-field isolation to 91 dB/\allowbreak{}mm. Our demonstration of field-free, stable, and programmable magnetoacoustic nonreciprocity provides a viable route toward miniaturized and highly integrated passive RF isolators and circulators~\cite{ref39}. More broadly, the strategy of using exchange bias to introduce built-in symmetry breaking and precisely engineer the spin-wave dispersion can be extended to a wide range of magnetoacoustic devices, including SAW-based magnetic sensors and tunable SAW filters.

% Source Word paragraph 52
\par
\vspace{6pt}

\begin{acknowledgments}
\textit{Acknowledgments---}This work was supported by the National Natural Science Foundation of China (Grant Nos. 12241404, 52225106, T2394471, 52631009, and 52561160112), the Natural Science Foundation of Beijing Municipality (Grant No. F251037), the Tsinghua University Initiative Scientific Research Program, and the New Cornerstone Science Foundation through the XPLORER PRIZE.
\end{acknowledgments}

% Manual bibliography: suppress REVTeX's automatic BibTeX lookup.
\makeatletter
\let\auto@bib@innerbib\@empty
\makeatother
% PRL references with article titles; numbering follows first citation.
% Full titles, authors and DOI metadata are retained in reference_metadata.json.

\end{document}